\documentstyle[12pt]{article}

\begin{document}


{\Large \bf Some $p$-adic differential equations }

\bigskip

\date{}

\bigskip

\noindent {{\large Maurice de GOSSON}   University of
Karlskrona-Ronneby, 371 79 Karlskrona, Sweden}

\bigskip

\noindent {{\large   Branko  DRAGOVICH}   Steklov Mathematical
Institute, Gubkin St. 8, 117966, Moscow, Russia;  Institute of
Physics, P.O.Box 57, 11001 Belgrade, Yugoslavia}

\bigskip

\noindent {{\large Andrei KHRENNIKOV}  Department of Mathematics,
Statistics and Computer Sciences, V$\ddot{a}$xj$\ddot{o}$
University, V$\ddot{a}$xj$\ddot{o}$, S-35195, Sweden}

\bigskip

\bigskip

\bigskip

\bigskip

\begin{abstract}
We investigate various properties of $p$-adic differential equations
which have as a solution an analytic function of the form
$$
F_k (x) = \sum_{n\geq 0}n! P_k(n)x^n ,
$$
where $P_k(n) = n^k + C_{k-1}n^{k-1} + \cdots + C_0 $ is a polynomial
in $n$ with $C_i \in {\bf Z}$ (in a more general case 
$C_i \in {\bf Q}$ or
$C_i \in {\bf C_p}$) , and the
region of convergence is $\mid x \mid_p
< p^{\frac{1}{p-1}}$. For some special classes of $P_k(n)$, as well as
for the general case, the existence
of the corresponding linear differential equations of the first-  and
second-order for $F_k(x)$, is shown. In some cases such equations
are constructed.
For the second-order differential equations there is no other
analytic solution of the form $\sum a_n x^n$.
Due to the fact that  the corresponding inhomogeneous
first-order differential equation exists one can construct 
infinitely many
inhomogeneous second-order equations with the same analytic solution.
Relation to some rational sums with the Bernoulli numbers and to
$F_k(x)$ for some $x\in {\bf Z}$ is considered. Some of these differential
equations can be related to $p$-adic dynamics and $p$-adic
information theory.
\end{abstract}

\section{Introduction}

Some aspects of the $p$-adic series of the form
$$
F_k (x) = \sum_{n\geq 0}n! P_k(n)x^n ,           \eqno(1.1)
$$
where $P_k(n) = n^k + C_{k-1}n^{k-1} + \cdots + C_0 $ is a polynomial
in $n$ with $C_i \in {\bf Z}$, have been considered in 
a few of articles (see [1],
[2] and references therein). It was noted in [1] that
$$
  F_0(x) = \sum_{n\geq 0}n!x^n                     \eqno(1.2)
$$
is an analytic solution of the following $p$-adic differential
equation:
$$
  x^2 w^{''}(x) + (3x-1) w'(x) + w(x) = 0.        \eqno(1.3)
$$

Here we investigate the existence, construction and various properties
of the differential equations which have as an analytic solution
$p$-adic power series of the form (1.1) with
$$
 P_k(n) = n^k + C_{k-1}n^{k-1} + \cdots + C_0 , \ \  C_i \in {\bf Q} .    \eqno(1.4)
$$
In a sense we mainly consider
an inverse problem related to differential equations, i.e. we
are looking for differential equation for which a solution is known.

Recall that the power series (1.1) has $p$-adic region of convergence
$ D_p = \{x \in {\bf C_p} : \mid x \mid_p < p^{\frac{1}{p-1}}
   \}$, where
${\bf C_p}$ is the algebraic closure of ${\bf Q_p}$ [3].
In the case of restriction to ${\bf Q_p}$, 
we have $D_p = {\bf Z_p}$ for every $p$.
Note that in the real case the series (1.1) is not convergent for
any $0\neq x\in {\bf Q}$.

A theory of the $p$-adic hypergeometric differential equations 
is presented in Dwork's book [4].

\bigskip

\bigskip

\section{Existence of some $p$-adic differential equations}

It is not difficult to verify that expression (1.2), which
is the simplest example of (1.1), satisfies not only equation
(1.3) but also the first-order inhomogeneous differential
equation
$$
  x^2 w' + (x-1)w = -1.                       \eqno(2.1)
$$
Note that differentiation of (2.1) gives (1.3).

Combining (1.3) and (2.1) in the form
$$
   x^2 w^{''} + (3x-1)w' + w + R(x)[x^2 w' + (x-1)w +1] = 0,
\eqno(2.2)
$$
where $R(x)$ is a rational function 
with integer coefficients, one can consider infinitely many
second-order linear inhomogeneous $p$-adic differential equations
with the same analytic solution (1.2). Generally, we will be interested
in differential equations of the form
$$
   (Polynomial)_1 w^{''} + (Polynomial)_2 w' + (Polynomial)_3 w
$$
$$
= (Polynomial)_4 ,
\eqno(2.3)
$$
where the polynomials are in $x$ with integer (or
$p$-adic) coefficients, and $w = F_k(x)$ with $P_k(n)$ given by
(1.4).

\bigskip

{\bf Proposition 1}  Let $A(x)$ and $B(x)$ be  rational
functions  with rational coefficients. If there are
differential equations
$$
A(x) F'_{\nu}(x) + B(x) F_{\nu}(x) = C ,   \ \  
C \in {\bf Q} , \eqno(2.4)
$$
$$
A(x) F^{''}_{\nu}(x) + (A'(x) + B(x)) F'_{\nu}(x) +
B'(x) F_{\nu}(x) = 0 ,         \ \                    \eqno(2.5)
$$
with the analytic solution
$$
  F_{\nu}(x) = \sum_{n\geq 0} n! P_{\nu}(n) x^n ,      \eqno(2.6)
$$
then there exist also similar differential equations of
the first- and second-order with the solution
$$
  F_{\mu +\nu}(x) = \sum_{n\geq 0} n! \prod_{i=1}^{\mu} (n+i)^2
  P_{\nu}(n+\mu)x^n .                       \eqno(2.7)
$$

{\it Proof}: Rewriting eq. (2.4) in the form
$$
\frac{A(x)}{B(x)} F'_{\nu}(x) + F_{\nu}(x) = \frac{C}{B(x)}
$$
and taking its derivative one obtains a new equation
$$
  A_1 (x) F^{''}_{\nu}(x) + B_1(x) F'_{\nu}(x) = C ,
$$
which is of the same form as (2.4) but with new rational functions
$A_1(x)$ and $B_1(x)$: $A_1(x) = -A(x)B(x)/B'(x),\  B_1(x)
= (B'(x)A(x) - A'(x)B(x) - B^2(x))/B'(x)$.
Repeating this procedure $\mu$ times, we get
$$
A_{\mu}(x) F_{\nu}^{(\mu +1)}(x) + B_{\mu}(x) F_{\nu}^{(\mu )} = C .
$$
Taking into account that $F_{\nu}^{(\mu)}(x) = F_{\mu + \nu}(x)$,
for functions $F_{\nu} (x)$ and $F_{\mu + \nu}(x)$ given by
(2.6) and (2.7), respectively, we have differential equation
for $F_{\mu + \nu}$:
$$
  A_{\mu}(x) F'_{\mu +\nu}(x) + B_{\mu}(x) F_{\mu + \nu}(x) = C ,
       \eqno(2.8)
$$
which resembles equation (2.4). The corresponding second-order differential
equation is
$$
A_{\mu}(x) F^{''}_{\mu +\nu}(x) + (A'_{\mu}(x) + B_{\mu}(x))
F'_{\mu +\nu}(x) + B'_{\mu}(x) F_{\mu +\nu}(x) = 0.         \eqno(2.9)
$$

\bigskip

From the proof of the Proposition 1 it also follows

\bigskip

{\bf Corollary 1} Derivatives of any order of the function
(2.6), which is related to equations (2.4)and (2.5), 
induce the corresponding
first- and second-order differential equations.

\bigskip

{\bf Proposition 2} If there are differential equations (2.4) and (2.5)
 with the analytic solution (2.6), then there exist also similar differential equations
with the analytic solution
$$
G_{\nu}(x) = x^{m} F_{\nu}(x) =   x^m \sum_{n\geq 0} n! P_{\nu}(n) x^n , \ \
m \in {\bf N} .                        \eqno(2.10)
$$

{\it Proof}: Differentiating (2.10), and replacing $F_{\nu}(x)$
and $F'_{\nu}(x)$ in (2.4) one gets similar equation
$$
  A_1(x) G'_\nu (x) + B_1(x) G_\nu (x) = C ,       \eqno(2.11)
$$
where $A_1(x) = A(x)/x^m$ and $B_1(x) = B(x)/x^m - mA(x)/x^{m+1}$.
By differentiation of (2.11) one has the corresponding
second-order differential equation.

\bigskip

{\bf Proposition 3} There exist the first- and second-order
differential equations with the analytic solution
$$
  F_k (x) = \sum_{n\geq 0} n! n^k x^n, \ \ k=1,2,...     \eqno(2.12)
$$

{\it Proof}: Start with (1.2) which induces equations (2.1)
and (2.2). According to the Corollary 1, $F'(x) = \sum n! n x^{n-1}$
has its own differential equation. Due to the Proposition 2 it
follows that there exist equations for $F_1(x) = xF'(x)= \sum n!
n x^n$. Performing this procedure $k$ times we come to the Proposition 3.

\bigskip

{\bf Proposition 4} There exist a first- and a 
second-order differential
equation with the analytic solution
$$
\Phi_\alpha (x) = \sum_{n\geq 0} n! (n+ \alpha )x^n ,  
\ \ \alpha \in {\bf Q} .
\eqno(2.13)
$$

{\it Proof}: Let us introduce $G_\alpha (x) = x^\alpha F_0 (x) =
\sum_{n\geq 0} n! x^{n+\alpha} $. According to the Proposition 2,
$G_\alpha (x)$ is an analytic solution of a first- 
and second-order differential
equation if $\alpha \in {\bf N}$. In the same way one can show that
$G_\alpha (x)$ is also a solution of a first-order 
differential equation
if $\alpha \in {\bf Q}$, as well as if $\alpha \in {\bf C_p}$.
Differentiating equation for $G_\alpha (x)$ in an appropriate way
one can obtain the corresponding first-order differential equation
for $G'_\alpha (x)$ (see also Corollary 1). In an analogous way
to the Proposition 2 it follows that $\Phi_\alpha (x) = x^{-\alpha}
G'_\alpha (x) =\sum_{n\geq 0}n! (n+ \alpha)x^n $ is an analytic
solution of some first- and second-order  differential equations.

\bigskip

It is now obvious that any $p$-adic power series of the form
$$
F_k (x)= \sum_{n\geq 0}n! \prod_{i=1}^l (n+\alpha_i)^{k_i}
x^n , \ \ k_1 + k_2 +\cdots + k_l =k, \ \ \alpha_i \in {\bf Q} ,
                               \eqno(2.14)
$$
is an analytic solution of a first- and, consequently,
of a second-order  homogeneous differential equation.

We can take that in (2.14) some or all of $\alpha_i \in {\bf Q_p}$
(or ${\bf C_p}$), but in such case there is restriction of
our consideration to a definite ${\bf Q_p}$ (or ${\bf C_p}$). However,
taking $\alpha_i \in {\bf Q}$ we have results valid in ${\bf C_p}$
for every $p$.

\bigskip

{\bf Theorem 1} To each function of the form $F_k =
\sum_{n\geq 0}n! P_k(n) x^n$, where $P_k(n) = n^k + C_{k-1}
n^{k-1} +\cdots + C_0$ is a polynomial in $n$ with coefficients
$C_i \in {\bf Q}$ (or $C_i \in {\bf C_p}$), corresponds a first-order
differential equation, and consequently the second-order one.

{\it Proof}: It follows from  the fact that
the above polynomial
$P_k (n)$ can be rewritten in the form
$$
P_k (n) = \prod_{i=1}^k (n+\alpha_i ),
$$
where $\alpha_i \in {\bf C_p}$.

\bigskip

\bigskip

\section{Construction of some $p$-adic differential equations}

There are many ways to construct relevant differential
equations for some $F_k(x)=\sum n! P_k (n) x^n$ with simple
polynomials $P_k (n)$.

For functions  $\sum n! n^k x^n$, where $k=0,1,2,\cdots$,
the relations  [1] of the following form are valid:
$$
 x^k\sum n! n^k x^n + U_k (x)\sum n! x^n = V_{k-1}(x),   \eqno(3.1)
$$
where $U_k(x)$ and $V_{k-1}(x)$ are certain polynomials in $x$
with integer coefficients. The first three of them are:
$$
  x\sum_{n\geq 0} n! nx^n + (x-1)\sum_{n\geq 0} n! x^n = -1 ,     \eqno(3.2)
$$
$$
  x^2\sum_{n\geq 0} n! n^2 x^n  +  (-x^2 +3x -1)\sum_{n\geq 0}
n!x^n = 2x-1 ,  \eqno(3.3)
$$
$$
x^3\sum_{n\geq 0} n!n^3 x^n + (x^3 -7x^2 +6x -1)\sum_{n\geq 0} n!x^n =
-3x^2 + 5x -1 .                                    \eqno(3.4)
$$
We use the above  relations for power series  to construct differential
equations for some simple cases of $F_k (x)$.

\bigskip

{\bf Example 1: $F_0(x) = \sum n! x^n$}.

Starting with (3.2) one obtains
$$
  x^2 F'_0(x) + (x-1) F_0(x) = -1 ,             \eqno(3.5)
$$
that is the equation (2.1). Differentiation of (3.5) gives
$$
 x^2 F^{''}_0 (x) + (3x-1) F'_0 (x) +F_0(x) = 0 ,  \eqno(3.6)
$$
which is just (1.3).

\bigskip

{\bf Example 2: $F_1(x) = \sum n! n x^n$}.

Due to (3.2) and (3.3) one gets
$$
  x^2(x-1) F'_1(x) + (x^2 - 3x +1) F_1(x) = x .     \eqno(3.7)
$$
Dividing (3.7) by $x$ and performing derivation one has
$$
  x^3 F^{''}_1 (x) + x(3x-1) F'_1 (x) + (x+1) F_1 (x) = 0.
                                                   \eqno(3.8)
$$

\bigskip

{\bf Example 3: $F_1(x) = \sum n!(n+1) x^n$}.

Combining (3.2) and (3.3) we obtain:
$$
  x^2 F'_1(x) + (2x-1)F_1(x) = -1 ,       \eqno(3.9)
$$
$$
  x^2 F^{''}_1(x) + (4x-1)F'_1(x) + 2F_1(x) = 0.  \eqno(3.10)
$$

\bigskip

{\bf Example 4: $F_2(x) = \sum n! (n+1)(n+2)x^n$}.

The corresponding differential equations are:
$$
 x^2 F'_2(x) + (3x -1) F_2 (x) = -2 ,        \eqno(3.11)
$$
$$
 x^2 F^{''}_2(x) + (5x-1) F'_2(x) + 3F_2(x) = 0 , \eqno(3.12)
$$
and can be obtained using equations (3.2), (3.3) and (3.4).

Examples 3 and 4 are particular ones of $F_1(x) =
\sum n! (n+ \alpha) x^n , \  \alpha \in {\bf Q}$. Let us
construct now the corresponding differential equations
for any $\alpha \in {\bf Q}$, which does exist according to the
Proposition 4.

\bigskip

{\bf Example 5:  $\Phi_\alpha (x) = \sum n! (n+\alpha) x^n ,\ 
\alpha\in {\bf C_p} $}.

It is worthwhile to start with $G_\alpha (x) = x^{\alpha}
F_0(x) = \sum n! x^{n +\alpha}$. Substituting $F_0 (x)
= x^{-\alpha}G_\alpha (x)$ in its equation (3.5) one obtains
$$
  x^2 G'_\alpha (x) - [(\alpha - 1)x + 1] G_\alpha (x)
= -x^{\alpha} .
$$
Forming the second-order differential equation for $G_\alpha (x)$
and taking $G'_\alpha (x) = x^{\alpha - 1} \Phi_\alpha (x)$
we have
$$
  x^2 [(\alpha - 1)x + 1]\Phi'_\alpha (x) +
[(\alpha - 1)x^2 - (\alpha - 3)x - 1]\Phi_\alpha (x) =
-(\alpha - 1)^2 x - \alpha ,                     \eqno(3.13)
$$
and consequently
$$
  x^2 [(\alpha -1)x +1][(\alpha -1)^2 x + \alpha] \Phi^{''}_\alpha (x)
$$
$$
+ [3(\alpha -1)^3 x^3  - (\alpha -1)(\alpha^2 - 9\alpha +4)x^2
  -(2\alpha^2 -7\alpha +1)x - \alpha] \Phi'_\alpha (x)
$$
$$
+[(\alpha -1)^3 x^2 + 2\alpha (\alpha -1)x + (\alpha +1)] \Phi_\alpha (x)
= 0 .                                               \eqno(3.14)
$$

\bigskip

{\bf Example 6: $F_k(x) = \sum n! \prod_{i=1}^k (n+i)x^n , \ 
k\in {\bf N}$}.

Starting from the Example 3,  using the method of mathematical
induction and an analogous way to the Example 5, one can derive the
following equations:
$$
  x^2 F'_k (x) + [(k+1)x -1] F_k(x) = -k! ,         \eqno(3.15)
$$
$$
  x^2 F^{''}_k(x) + [(k+3)x - 1] F'_k(x) + (k+1)F_k(x) = 0. \eqno(3.16)
$$
Note that (3.15) and (3.16) hold for $k=0$ as well.

\bigskip

{\bf Example 7:  $F_{k}(x) = \sum n! \prod_{i=1}^k (n+i)^2 x^n$}.

Differentiation of equation (3.5) $k$ times yields:
$$
  x^2 F^{(k+2)}_0(x)  + [(2k +3)x - 1]F^{(k+1)}_0(x)  + (k+1)^2
F^{(k)}_0(x) = 0 .
$$
Since $F^{(k)}_0(x) = F_k(x)$ we have
$$
  x^2 F^{''}_k(x) + [(2k+3)x -1]F'_k(x) + (k+1)^2F_k(x) = 0 . \eqno(3.17)
$$

The corresponding first-order equation of (3.17) has a rather complex form.

\bigskip

{\bf Example 8: $\Phi_{\alpha\beta}(x) = 
\sum n! (n+\alpha)(n+\beta) x^n ,
\ \ \alpha ,\beta \in {\bf C_p}$}.

Denote $G_{\alpha\beta}(x) = x^{\beta}\Phi_\alpha (x)= 
\sum n!(n+\alpha)
x^{n+\beta}$ and note that $G^{'}_{\alpha\beta}=x^{\beta - 1}
\Phi_{\alpha\beta}(x)$. Using equation (3.13) for $\Phi_\alpha(x)$  one
can obtain the following differential equation:
$$
 x^2 [(\alpha -1)x +1][(\alpha -1)(\beta - 1)x^2 + (\alpha +\beta -3)x + 1]
 \Phi'_{\alpha\beta}(x)
$$
$$
 +  \{ x(\beta -1)[(\alpha -1)x +1][(\alpha -1)
(\beta -1)x^2 + (\alpha +\beta -3)x +1]
$$
$$
 +x[3x(\alpha -1)+2]
[(\alpha -1)(\beta -1)x^2 + (\alpha +\beta -3)x+1]
$$
$$
 - x^2 [(\alpha -1)x +1]
 [(\alpha - 1)(\beta -1)2x +\alpha +\beta -3] $$ $$
 -[(\alpha -1)(\beta -1)x^2 +(\alpha +\beta -3)x +1]^2  \}
 \Phi_{\alpha\beta}(x)
$$
$$
 = x [(\alpha -1)^2 x +\alpha]
 [2(\alpha -1)(\beta -1)x +\alpha +\beta -3] -[(\alpha
-1)^2 (\beta +1)x + \alpha\beta]
$$
$$
\times[(\alpha -1)(\beta
-1)x^2 + (\alpha +\beta -3)x +1]  . \eqno(3.18)
$$

The corresponding homogeneous second-order differential equation
exists, but it is more complex than (3.18).

\bigskip

{\bf Example 9: $F_2(x) = \sum n! n^2 x^n$}.

This can be considered as special case of the Example 9 for
$\alpha =\beta =0$. From (3.18) it follows
$$
 x^2 (x^2 - 3x +1) F'_2(x) + (x^3 - 7x^2 + 6x -1)
 F_2(x) = - x(x+1).                               \eqno(3.19)
$$
The corresponding homogeneous second-order differential equation is
$$
  x^3 (x+1)(x^2 -3x +1) F^{''}_2(x) + x(3x^4 - 6x^3 -7x^2 +6x -1)
  F'_2(x)
$$
$$
  + (x^4 +2x^3 - 13x^2 + 2x +1) F_2(x) = 0.    \eqno(3.20)
$$

It is obvious that using the above procedures one can construct
differential equation for any function of the form
$F_k(x) = \sum n! \prod_{i=1}^k (n+\alpha_i) x^n$, where 
$\alpha_i \in {\bf C_p}$.

\bigskip

\bigskip

\section{On other solutions}

It seems that the homogeneous second-order differential
equation for analytic function $F_k(x) = \sum n! P_k (n) x^n$ has
not another analytic solution in the region containing point $x=0$.
Namely, in any particular case of the above examples one
can start with power series expansion and conclude that only
$F_k(x) = \sum n! P_k(n) x^n$ is the corresponding analytic solution.
However, the corresponding general statement needs a clear rigorous
proof.

Note that the solution $F_0(x) = \sum_{n \geq 0} n! x^n$ 
can be presented in the form
$$
  F_0(x) = \sum_{n\geq 0} b_n (x -\beta)^n ,      \eqno(4.1)
$$
where coefficients $b_n$ satisfy conditions
$$
  \sum_{n\geq k} b_n \left( \begin{array}{c} n \\  k
  \end{array} \right) \beta^{n-k} = k! , \ \  k=0,1,2,...
                                                  \eqno(4.2)
$$
Solution of the system of equations (4.2) yields $$
 b_n = \sum_{k\geq n} (-1)^{k-n} k! \left(
 \begin{array}{c} k \\ n \end{array} \right) \beta^{k-n} .
                                                  \eqno(4.3)
$$

One can easily verify that in the simplest case, given by
the Example 1 and equation (1.3), one has the following
two new solutions (see also [5]):
$$
 w_1(x) = \frac{1}{x}\exp{(-\frac{1}{x})} , \ \
 w_2(x) =
\frac{1}{x}\exp{(-\frac{1}{x})} \int^x_{x_0}
\frac{1}{t}\exp{(\frac{1}{t})}dt ,
                                                   \eqno(4.4)
$$
where the region of $p$-adic convergence of $w_1(x)$ and $w_2(x)$
in (4.4) is ${\Delta }_p = \{ x \in {\bf C_p} :
\mid x \mid_p > p^{\frac{1}{p-1}} \}$.
Thus ${\bf C_p} = D_p \cup S_p \cup {\Delta}_p$, where $D_p$ is
the region of convergence
of analytic solution (1.2) and $S_p$ is the sphere
$ S_p = \{x \in {\bf C_p} : \mid x \mid_p = p^{\frac{1}{p-1}} \}$.
Note that $D_p,\  S_p, $ and $ {\Delta}_p $ are mutually
disjoint subsets of ${\bf C_p}$.

Using a reasoning analogous to the preceding section, one can show
that all homogeneous second-order differential equations
for $F_k(x) = \sum n! P_k(n) x^n $
have the corresponding two other solutions which are connected
with (4.4) in the similar way as analytic solutions $F_k(x)$
are related to $F_0(x)$.

\bigskip

\bigskip

\section{Relation to rational summation of $p$-adic series}

The above differential equations may be used to obtain various
expressions for sums of some $p$-adic series.

For example, from (3.15) one can rederive (3.2)-(3.4), as well as
any other sum of the form
$$
  \sum n! [n^k + u_k (x)]x^n = v_k (x),            \eqno(5.1)
$$
where $u_k (x)$ and $v_k (x)$ are rational functions of variable
$x$. Any other possible rational sum can be generated from (5.1)
multiplying it by rational numbers and performing the corresponding
summation. For $k=1,...,5$ we calculated (5.1) in the explicit
form:
$$
   \sum_{n\geq 0} n!\left( n + \frac{x-1}{x}\right) x^n = \frac{-1}{x},
                                                         \eqno(5.2)
$$
$$
     \sum_{n\geq 0} n!\left( n^2 + \frac{-x^2+3x-1}{x^2}\right) x^n
= \frac{2x-1}{x^2},                                  \eqno(5.3)
$$
$$
     \sum_{n\geq 0} n!\left( n^3 + \frac{x^3 - 7x^2 +6x-1}{x^3}\right) x^n
= \frac{-3x^2 +5x-1}{x^3},                             \eqno(5.4)
$$
$$
   \sum_{n\geq 0} n!\left( n^4 + \frac{-x^4 +15x^3 - 25x^2 +10x-1}{x^4}
\right) x^n
$$
$$
= \frac{4x^3- 17x^2 +9x-1}{x^4}, \eqno(5.5)
$$
$$
    \sum_{n\geq 0} n!\left( n^5 + \frac{x^5-31x^4 +90x^3-65x^2+15x-1}
{x^5}\right) x^n
$$
$$
 = \frac{-5x^4+49x^3-52x^2+14x-1}{x^5}.                  \eqno(5.6)
$$

Taking $x=t \in {\bf Z}$ in (5.2)-(5.6) we obtain $p$-adic sums valid
in all ${\bf Q_p}$. The case $x=1$ and $k=1,...,11$ is presented in [1].
For some evaluation of $\sum n!$ one can see Schikhof's book ([3], p. 17).
We write down sums for $x= -1$ and   $k= 1,...,5$:
$$
   \sum_{n\geq 0} (-1)^n n! (n+2) = 1, \ \ \
   \sum_{n\geq 0} (-1)^n n! (n^2-5) = -3, \ \ \
$$
$$
   \sum_{n\geq 0} (-1)^n n! (n^3+15) = 9, \ \ \
   \sum_{n\geq 0} (-1)^n n! (n^4-52) = -31, \ \ \
$$
$$
\sum_{n\geq 0} (-1)^n n! (n^5+203) = 121. \ \ \
                                                      \eqno(5.7)
$$

Note also that putting $x=1/(1-\alpha), \ \ x=1$ and $x=-1$
successively in (3.13) we have:
$$
 \sum_{n\geq 0} n! (n+\alpha)\left( \frac{1}{1-\alpha} \right)^n
= \alpha -1, \ \ \ \mid 1-\alpha\mid_p^{-1} < p^{\frac{1}{p-1}},
                                                     \eqno(5.8)
$$
$$
 \sum_{n\geq 0} n! (n+\alpha)(\alpha n +1)
= -\alpha^2 +\alpha -1,
                                                     \eqno(5.9)
$$
$$
 \sum_{n\geq 0} (-1)^n n! (n +\alpha)[(\alpha -2)n +2\alpha -5]
= \alpha^2 -3\alpha +1.
                                                     \eqno(5.10)
$$
The sum (5.10) can be easily verified employing (5.7).

Since the $p$-adic sums (5.2)-(5.6) are convergent in ${\bf Z_p}$ one
can use them to obtain a new kind of $p$-adic sums with the
Bernoulli numbers $B_n$, which may be regarded as [3]
$$
   B_n = \int_{Z_p} x^n dx,   \ \ \ n=0,1,2,...,
$$
where $\int_{Z_p} f(x) dx$ denotes the Volkenborn integral.
Recall that expressions
$$
  B_0 =1,\ \ \ \ \sum_{i=1}^{n-1}{n\choose i} B_i = 0,
\ \ \  n\geq 2
$$
determine all Bernoulli numbers.
Rewriting (5.2)-(5.6) in the form (3.1) and performing the
Volkenborn integration, we get the first five sums:
$$
 \sum_{n\geq 0} n! [(n+1)B_{n+1}- B_n] = -1,
$$
$$
 \sum_{n\geq 0} n! [(n^2-1)B_{n+2} +3B_{n+1}- B_n] = -2,
$$
$$
 \sum_{n\geq 0} n! [(n^3+1)B_{n+3}-7B_{n+2}+6B_{n+1}- B_n] = -4,
$$
$$
 \sum_{n\geq 0} n! [(n^4-1)B_{n+4}+15B_{n+3}-25B_{n+2}+10B_{n+1}- B_n] = -
\frac{25}{3},
$$
$$
 \sum_{n\geq 0} n! [(n^5+1)B_{n+5}-31B_{n+4}+90B_{n+3}-65B_{n+2}
+15B_{n+1}- B_n] = -\frac{33}{2}. 
$$ 
The termwise integration of an analytic function is provided by 
the Proposition 55.4 of [3].
If we first make
transformation $x\to -x$ and then apply the Volkenborn integral we
can obtain the corresponding sums with $(-1)^n$ factors. As an
illustration we give the following two sums: $$ \sum_{n\geq 0}
(-1)^n n! [(n+1)B_{n+1}+ B_n] = 1, \ \ $$ $$ \sum_{n\geq 0} (-1)^n
n! [(n^2-1)B_{n+2}-3B_{n+1}- B_n] = -2.
$$
Since $\mid B_n \mid_p
\leq p$ (see [3], p. 172), there are no problems with the
convergence of the above series in ${\bf Q_p}$ for 
every $p$ and results are valid in all ${\bf Q_p}$.
Multiplying the series (3.1) by $x^m$ before integration,
one can generalize the above formulas involving the Bernoulli numbers.

\bigskip

\bigskip

\section{Possible physical applications }

Since 1987, when a notion of  $p$-adic strings [6]
was introduced for the first time,
there have been exciting investigations in application of
$p$-adic numbers in many parts of modern theoretical and mathematical
physics (for a review, see, e.g. Refs. [7],[8] and [9]).
One of the very perspective approaches is related to adeles [10],
which unify $p$-adic and real numbers. So, adelic quantum theory
(see [11]-[13]) seems to be a more complete theory then the ordinary
one based on real and complex numbers only.

Some of the above $p$-adic differential equations may be regarded
as classical equations of motion in the Lagrangian formalism.
Recall that for a given Lagrangian $L(\dot{q},q,t)$, the classical
equation of motion is the Euler-Lagrange equation
$$
  \frac{\partial L}{\partial q} -\frac{d}{dt}\frac{\partial L}
{\partial {\dot{q}}} = 0 ,                             \eqno(6.1)
$$
where $\dot{q}$ denotes derivative of $q$ with respect to
the time variable $t$. In the case of quadratic Lagrangians, i.e.
$$
  L(\dot{q},q,t) = a(t){\dot{q}}^2 + 2b(t)\dot{q}q + c(t)q^2
+2d(t)\dot{q} + 2e(t)q + f(t),                    \eqno(6.2)
$$
the classical equation of motion reads:
$$
  a(t)\ddot{q} + \dot{a}(t)\dot{q}(t) + [\dot{b}(t) - c(t)]q(t)
= e(t) - \dot{d}(t) .                         \eqno(6.3)
$$

Let us consider the simplest case of our $p$-adic differential
equations presented in the form (2.2), where $R(x)$ is a rational
function with integer coefficients. According to (6.3), a second-order
differential equation can be  an equation of motion if there is a
definite relation between the coefficients of the terms with
$\ddot{q}$ and $\dot{q}$. One can easily see that the case
$R(x) =0$ does not lead to an equation of motion. However, if
$R(x) = (-x +1)/x^2 $ then (2.2) becomes equation of motion
in the following form:
$$
   t^4 \ddot{q} + 2t^3 \dot{q} + (2t-1)q = t-1 .   \eqno(6.4)
$$
One of the possible Lagrangians which give (6.4) is
$$
 L(\dot{q},q,t) = \frac{t^2}{2}{\dot{q}}^2 + \left(
 \frac{t^3}{3} + 2\log{t} + \frac{1}{t} + C \right) \dot{q}q
+ \frac{t^2}{2} q^2 - \frac{1}{t}\dot{q} + \frac{1}{t} q ,
                                                  \eqno(6.5)
$$
where $C$ is a constant. Other Lagrangians, which lead to
(6.4), have less symmetric coefficients than (6.5).
A solution of (6.4) is $q(t) = \sum n!t^n$ and represents
$p$-adic classical trajectory. In virtue of the Proposition 5
this is a unique $p$-adic analytic solution around $x=0$
and there is not the corresponding real analytic counterpart.

It is worth noting also that other analytic solutions of the form
$F_k(t) = \sum n! P_k(n)x^n$
have no real counterparts and may describe some dynamical systems
for which real numbers are useless. As a possible application of these
analytic solutions one can consider  dynamics on information spaces
introduced in [14].

\bigskip

\bigskip

\section{Concluding remarks}

When the coefficients $C_i$ of the polynomials $P_k(n)$ in (1.1) are
rational numbers and $x\in {\bf Q_p}$ then all the 
obtained results for
$F_k (x)$ are valid in ${\bf Z_p}$ for every $p$. Taking into account
solutions (see Section 4) which have real counterparts,
we can construct also  some  adelic [10] solutions. Namely, an adelic
solution for the case $k=1$ in the form (6.4) is:
$$
 q_(t) = ( q_\infty (t_\infty), F_0 (t_2), F_0 (t_3), ...,
  F_0 (x_p), ...),                                 \eqno(7.1)
$$
where the index $\infty$ denotes real case, $F_0 (t) =\sum n! t^n$,
and
$$
 q_\infty (t) =  \frac{1}{t}\exp{(-1/t)}
$$
$$
\times \left( A_1 + A_2
\int_{t_0}^t \exp{(2/y)}dy
+ \int_{t_0}^t dy\exp{(2/y)}
\int_{y_0}^y dz \frac{z-1}{z^3} \exp{(-1/z)} \right) ,  \eqno(7.2)
$$
where $A_1$ and $A_2$ are arbitrary integration constants.

All the above considered differential equations are linear.
Some of them are homogeneous and the others are inhomogeneous.
Rewriting all equations in the form
$$
 {\cal D}_k (\frac{d^2}{dx^2}, \frac{d}{dx}, x)F_k(x) =  0,
                                                   \eqno(7.3)
$$
where operator ${\cal D}_k$ linearly depends on derivatives
$\frac{d^2}{dx^2}$ and $\frac{d}{dx}$,  one can construct
many non-linear differential equations taking various
products of ${\cal D}_k$. For example, according to
(3.5) and (3.9), we have
$$
 [x^2 u' + (x-1)u + 1][x^2 u' + (2x - 1)u +1] = 0
                                               \eqno(7.4)
$$
with solutions: $u_1(x)= \sum n! x^n , \ \ u_2(x) = \sum n!
(n+1) x^n $.

\bigskip

\noindent{\bf Acknowledgments}

A part of this article has been done during the 
visit of one of the authors
(B.D.) to the Department of Mathematics, University of
Karlskrona-Ronneby, Sweden, and to the Institute of Mathematics
and System Engineering,
V$\ddot{a}$xj$\ddot{o}$ University, Sweden, on the basis
of the research project of the Royal Academy of Science of Sweden
in collaboration with States of Former SU. The work of B.D. was
supported in part by RFFI grant 990100866.

\end{document}